\documentclass[letterpaper]{IEEEtran}

\usepackage[utf8]{inputenc}
\usepackage{graphicx}
\usepackage{stfloats}
\usepackage{amssymb}
\usepackage[cmex10]{amsmath}
\usepackage{color}
\usepackage{theorem}
\usepackage{pifont}
\usepackage{array}
\usepackage{hyperref}
\usepackage{cite}
\usepackage{url}
\usepackage{bbm}
\usepackage{relsize}
\usepackage[ruled,vlined,titlenumbered]{algorithm2e_old}

\newtheorem{theorem}{Theorem}
\newtheorem{definition}{Definition}
\newtheorem{lemma}{Lemma}

\newtheorem{remark}{Remark}

\newtheorem{example}{Example}

\DeclareMathOperator{\wt}{wt}
\DeclareMathOperator{\rank}{rank}

\DeclareMathOperator{\defi}{def}

\newcommand{\qed}{\hfill \mbox{\raggedright \rule{.07in}{.1in}}}

\newcommand{\defeq}{\overset{\defi}{=}}

\newcommand{\F}{\mathbb F}

\renewcommand{\a}{\mathbf a}
\renewcommand{\c}{\mathbf c}
\newcommand{\e}{\mathbf e}
\renewcommand{\r}{\mathbf r}

\renewcommand{\i}{\mathbf i}

\newcommand{\A}{\mathbf A}
\newcommand{\B}{\mathbf B}

\newcommand{\G}{\mathbf G}

\newcommand{\0}{\mathbf 0}

\newcommand{\code}[1]{\ensuremath{\mathcal{C}_{#1}}}%code
\newcommand{\dfree}{\ensuremath{d_{\mathrm{free}}}}%free distance
\newcommand{\m}[1]{\ensuremath{{m}\left(#1\right)}}%Hamming metric

\newcommand{\dr}[1]{\ensuremath{{d}_{#1}^{r}}}%designed extended row distance
\newcommand{\drdes}[1]{\ensuremath{\oldbar{{d}}_{#1}^{r}}}%designed extended row distance

\newcommand{\dc}[1]{\ensuremath{{d}_{#1}^{c}}}%designed extended column distance
\newcommand{\dcdes}[1]{\ensuremath{\overline{{d}}_{#1}^{c}}}%designed extended column distance
\newcommand{\drc}[1]{\ensuremath{{d}_{#1}^{rc}}}%designed extended reverse column distance
\newcommand{\drcdes}[1]{\ensuremath{\overline{{d}}_{#1}^{rc}}}%designed extended reverse column distance

\newcommand{\da}{\ensuremath{d_{\alpha}}}

\newcommand{\en}[1]{\ensuremath{{\epsilon_{#1}}}}%erasure nodes
\newcommand{\comp}{\ensuremath{C}}
\newcommand{\order}[1]{\ensuremath{\mathcal{O}\left(#1\right)}}

\newcommand{\BMD}[1]{\ensuremath{\mathsf{BMD(#1)}}}

\newcommand{\printalgo}[1]
{
\begin{center}
\scalebox{0.85}{
\begin{algorithm} [H]
 #1
\end{algorithm}
}
\end{center}
}
\newcommand{\oldbar}{\bar}
\renewcommand{\bar}{\overline}

\IEEEoverridecommandlockouts

\begin{document}

\title{Efficient Decoding of Partial Unit Memory Codes of Arbitrary Rate}

\author{\IEEEauthorblockN{Antonia Wachter-Zeh$^1$, Markus Stinner$^2$ and Martin Bossert$^1$}\\
\IEEEauthorblockA{$^1$ Institute of Communications Engineering, University of Ulm, Ulm, Germany}\\
$^2$ Institute for Communications Engineering, Technical University of Munich, Munich, Germany\\

\texttt{antonia.wachter@uni-ulm.de, markus.stinner@tum.de, martin.bossert@uni-ulm.de}\\
\thanks{This work was supported by the German Research Council "Deutsche Forschungsgemeinschaft" (DFG) under Grant No. Bo 867/21-1. 
%This work was done while Markus Stinner was with the University of Ulm.
}}
\maketitle

\begin{abstract}
Partial Unit Memory (PUM) codes are a special class of convolutional codes, which are often constructed by means of block codes. 
Decoding of PUM codes may take advantage of existing decoders for the block code. 
The Dettmar--Sorger algorithm is %a so-called Bounded Minimum Distance decoding algorithm 
an efficient decoding algorithm for PUM codes, but allows 
only low code rates. The same restriction holds for several known PUM code constructions.
%Known constructions and a so-called Bounded Minimum Distance decoding algorithm for PUM codes allow only low code rates. 
In this paper, an arbitrary-rate construction, the analysis of its distance parameters and a 
generalized decoding algorithm for PUM codes of arbitrary rate are provided. 
%The Dettmar--Sorger
%decoding algorithm is generalized, 
The correctness of the algorithm is proven and it is shown that its complexity %of decoding one block 
is cubic in the length.
\end{abstract}


\section{Introduction}
The algebraic description and the distance calculation of convolutional codes is often difficult. 
By means of \emph{block} codes, special \emph{convolutional} codes of memory $m=1$
can be constructed, which enable the estimation of the distance parameters. 
Moreover, the existing efficient block decoders can be taken into account in order to decode the convolutional code. 
There are constructions of these so-called \emph{Partial Unit Memory} (PUM) codes \cite{Lee_UnitMemory_1976,Lauer_PUM_1979} based on Reed--Solomon (RS) \cite{zyablov_sidorenko_pum,Pollara_FiniteStateCodes,Justesen_BDDecodingUM},
BCH \cite{DettmarShav_NewUMCodes,DettmarSorger_PUMBCH} and -- in rank metric -- Gabidulin \cite{WachterSidBossZyb-PUMGabidulin_ISIT2011,WachterSidBossZyb_PUMbasedGab} codes.
Decoding of these PUM codes uses the algebraic structure of the 
underlying RS, BCH or Gabidulin codes.

In \cite{DettmarSorger_BMDofUM}, Dettmar and Sorger constructed low-rate PUM codes and decoded them up to half the extended row distance. 
Such a decoder is called \emph{Bounded Minimum Distance} (BMD) decoder for convolutional codes. 
Winter \cite{Winter_UM_PhD_1998} gave first ideas of an arbitrary rate construction. 

In this contribution, we construct PUM codes of arbitrary rate, prove their distance properties
and generalize the Dettmar--Sorger algorithm 
to PUM codes of arbitrary rate. 
We prove the correctness of the decoding algorithm and show that the complexity is cubic in the length. 
To our knowledge, no other construction and efficient decoding of PUM codes of arbitrary rate exist. 
Due to space limitations, we consider only PUM codes, but all results apply also to Unit Memory codes.

This paper is organized as follows. 
In Section~\ref{sec:def}, we give basic definitions, Section~\ref{sec:constr} provides the arbitrary rate construction and 
calculates its parameters. In Section~\ref{sec:decoding}, we explain and prove the BMD decoding algorithm. Section~\ref{sec:concl} concludes this contribution. 

\section{Definitions and Notations}\label{sec:def}

%\subsection{Notations}

Let $q$ be a power of a prime and let $\F$ denote the finite field of order $q$. % and $\F =\Fqs$ its extension field of order $q^s$. 
%Throughout this paper, we use $\Fq^{s \times n}$ to denote the set of all $s\times n$ matrices over $\Fq$ and 
We denote by $\F^n =\F^{1 \times n}$ the set of all \emph{row} vectors of length $n$ over $\F$ and  
%Let us denote the row space of a matrix $\A$ over $\Fq$ by $\langle \A \rangle$ and let 
%$\mathbf I_{s}$ denote the $s \times s$ identity matrix.
the elements of a vector $\a_j \in \F^n$ by $\a_j = (a_0^{(j)}, a_1^{(j)}, \dots, a_{n-1}^{(j)})$.

%\subsection{Convolutional Codes}
Let us define a zero-forced terminated convolutional code $\mathcal C$ for some integer $L$ 
by the following $L k \times (n(L+m))$ generator matrix $\G$ over the finite field $\F$ %with the submatrices $\mathbf G_0, \mathbf G_1, \dots ,\G_m$, 
%\begin{small}
\begin{equation}\label{eq:genmat_termin}
\G = 
\begin{pmatrix}
\G_0 & \G_1 & \dots & \G_m &&& \\
&\G_0 & \G_1 & \dots & \G_m&& \\
&&\ddots&&&\ddots&\\
&&&\G_0 & \G_1 & \dots & \G_m
\end{pmatrix},
\end{equation}
%\end{small}
%Let us define a convolutional code by its semi-infinite generator matrix $\G$ over a finite field $\F$ with the submatrices $\mathbf G_0, \mathbf G_1, \dots ,\G_m$, 
where $\mathbf G_i$, $i=0,\dots,m$ are $k \times n$--matrices and  
$m$ denotes the memory of $\mathcal C$ as in \cite{Johannesson_Fund_Conv_Codes}. 
% Each convolutional code can be represented by a minimal trellis. 
In the following, $N \defeq L+m$.

The error-correcting capability of convolutional codes is determined by \emph{extended (or active) distances}. 

Let $\mathcal C^r(j)$ denote the set of all codewords corresponding to paths in the minimal code trellis 
that diverge from the zero state at depth $0$ and return to the zero state 
for the first time at depth $j$.
%at depth $j$, possibly ``touching'' the zero state only in non-consecutive zero states in between. 
The extended row distance of order $j$ is defined as the minimum Hamming weight of all codewords in $\mathcal C^r(j)$:
% \begin{definition}[Extended Row Rank Distance]\label{def:extrowdist}
% The extended row rank distance of order $j=1,2,\dots$ is defined as
\begin{equation*}%\label{eq:defextrowrank}
\dr{j} \defeq \min_{\substack{\mathbf c \in \mathcal C^r(j)}} \lbrace \wt(\c) \rbrace.
\end{equation*}
% \end{definition}
Similarly, let $\mathcal C^c(j)$ denote the set of all codewords leaving the zero state at depth $0$ and ending in any state at depth $j$ and let 
$\mathcal C^{rc}(j)$ denote the set of all codewords starting in any state at depth $0$ and ending in the zero state in depth $j$, 
both without 
%consecutive 
zero states in between. 
The extended column distance and the extended reverse column distance are:
\begin{equation*}
\dc{j} \defeq \min_{\substack{\mathbf c \in \mathcal C^c(j)}} \lbrace \wt(\c) \rbrace, \quad
\drc{j} \defeq \min_{\substack{\mathbf c \in \mathcal C^{rc}(j)}} \lbrace \wt(\c)\rbrace.
\end{equation*}
The \emph{free} distance is the minimum (Hamming) weight of any non-zero codeword of $\mathcal C$ and can be determined by
% \begin{equation*}
$\dfree = \min_{\substack{j}}\lbrace \dr{j}\rbrace$.
% \end{equation*}
The extended row distance $\dr{j}$ can be lower bounded by a linear function 
% $\dr{j} \geq \max \lbrace \alpha j +\beta,\dfree \rbrace$ where $\beta \leq \dfree$ and $\alpha$ denotes the \emph{slope}, which is defined by:
with \emph{slope} $\alpha$:
\begin{equation*}
\alpha = \lim_{\substack{j \rightarrow \infty}} \Big \lbrace \frac{\dr{j}}{j} \Big \rbrace.
\end{equation*}

%TODO: Warum diese Definition der extended distances?

% TODO: Def. convolutional code, Def. free distance, active distances
% bei memory, constraint length auf Johannesson verweisen 
%\subsection{(Partial) Unit Memory Codes}\label{subsec:pum}
PUM codes are convolutional codes of memory $m=1$. Therefore, the semi-infinite generator matrix consists 
of two $k\times n$ sub-matrices $\G_0$ and $\G_1$. 
% The semi-infinte generator matrix of a PUM code is
% \begin{equation}\label{eq:def_Gm1}
% \mathbf G = \left( \begin{array}{ccccc}
% \mathbf G_0 &\mathbf G_1&& \\
% &\mathbf G_0 &\mathbf G_1& \\
% &&\ddots&\ddots\\
%              \end{array}\right),
% \end{equation}
% where $\mathbf G_0$ and $\mathbf G_1$ are $k\times n$ matrices. 
Both matrices have full rank if we construct an $(n,k)$ UM code. 
For an $(n,k\;|\;k_1)$ PUM code, $\rank(\mathbf G_0)=k$ and $\rank(\mathbf G_1)=k_1 < k$ hold, such that:
\begin{equation}\label{eq:pum_submatrices}
\mathbf G_0 = \left( \begin{array}{c} \mathbf G_{00}\\ \mathbf G_{01}\end{array}\right), \qquad \mathbf G_1 = \left( \begin{array}{c} \mathbf G_{10}\\ \mathbf 0\end{array}\right),
\end{equation}
where $\mathbf G_{00}$ and $\mathbf G_{10}$ are $k_1 \times n$ matrices and $\mathbf G_{01}$ is a $(k-k_1)\times n$-matrix. 
The encoding rule for a code block of length $n$ is given by 
$\mathbf c_j = \mathbf i_j \cdot \mathbf G_0 + \mathbf i_{j-1} \cdot \mathbf G_1$, for $\i_j, \i_{j-1} \in \F^{k}$.
% The memory of PUM codes is $m=1$, the overall constraint length of UM codes is $\nu = k$ and of PUM codes $\nu = k_1$. 

The free distance of UM codes is upper bounded by $\dfree \leq 2n-k+1$ and of PUM codes by $\dfree \leq n-k +k_1+1$. For both the slope is upper bounded by $\alpha \leq n-k$  \cite{Pollara_FiniteStateCodes,Thommesen_Justesen_BoundsUM}.

As notation, let the generator matrices
\begin{equation*}
 \G_0 ,%=\left(\begin{matrix}\mathbf G_{00}\\\mathbf G_{01}\end{matrix}\right), 
 \ \left(\begin{matrix}\mathbf G_{01}\\\mathbf G_{10}\end{matrix}\right),
 \ \mathbf G_{01}
 \ \text{and} \ \mathbf G_{\alpha} = \left(\begin{matrix}\mathbf G_{00}\\\mathbf G_{01}\\\mathbf G_{10}\end{matrix}\right)
\end{equation*}
define the block codes \code{0}, \code{1}, \code{01} and \code{\alpha} with the minimum Hamming distances $d_0$, $d_1$, $d_{01}$ and $d_{\alpha}$
and the BMD \emph{block} decoders \BMD{\code{0}}, \BMD{\code{1}}, \BMD{\code{01}} and \BMD{\code{\alpha}}, which correct errors up to half their minimum distance.

\section{Constructing PUM Codes of Arbitrary Rate}\label{sec:constr}
\subsection{Construction}
Since each code block of length $n$ of the PUM code can be seen as a codeword of the block 
code $\code{\alpha}$, a great $\da$ is important for the distance parameters of the convolutional code as well 
as for the decoding capability. 
%Clearly, the highest possible $\da$ is achieved if $\G_{\alpha}$ defines a 
One approach is to define by $\G_{\alpha}$ a \emph{Maximum Distance Separable} (MDS) code and $\da = n-k-k_1+1$. 
This is basically the construction from \cite{DettmarShav_NewUMCodes,DettmarSorger_BMDofUM} which designs \emph{low-rate} PUM codes since the $(k + k_1) \times n$ matrix $\mathbf G_{\alpha}$
can define an MDS code only if $k + k_1 \leq n$. Otherwise (as observed by \cite{Winter_UM_PhD_1998}), there are linear dependencies between the rows of $\mathbf G_{\alpha}$, 
% The construction having linearly dependent rows 
what we have to consider when constructing PUM codes of arbitrary rate. In the following, we provide a construction 
of arbitrary $k_1 < k$ and calculate its distance parameters. 

Let $k+k_1-\varphi \leq n$, for some $\varphi < k_1$, and let the $(k+k_1-\varphi)\times n$ matrix 
\begin{equation}\label{eq:Gtot}
\!\G_{tot} = 
\left(\begin{matrix}
\mathbf A\\
\mathbf \Phi\\
\mathbf G_{01}\\
\mathbf B\\
\end{matrix}\right)
\;
\text{with the sub-sizes}
\ \
\begin{matrix}
\hspace{-1.5ex}\mathbf A: (k_1-\varphi) \times n \\
\hspace{-8.5ex}\mathbf \Phi: \varphi \times n\\
\mathbf G_{01}: (k-k_1)\times n\\
\hspace{-1.5ex}\mathbf B: (k_1-\varphi)\times n\\
\end{matrix}
\end{equation}
%have full rank. 
define an MDS (e.g. RS) code. 
We define the sub-matrices of the semi-infinite generator matrix of the PUM code as follows in order 
to enable arbitrary code rates.

\begin{definition}[PUM Code of Arbitrary Rate]\label{def:pumcodearbrate}
Let $k_1 < k < n$ and let $\G_{tot}$ be defined as in \eqref{eq:Gtot}. 
Then, we define the PUM code by the following submatrices \eqref{eq:pum_submatrices}:
\begin{equation}\label{eq:winter-generator}
\mathbf G_0 = 
\left(\begin{matrix}
\mathbf G_{00}\\
\mathbf G_{01}
\end{matrix}\right) =
\left(\begin{matrix}
\mathbf A\\
\mathbf \Phi\\
\mathbf G_{01}
\end{matrix}\right), 
\quad
\mathbf G_1 = 
\left(\begin{matrix}
\mathbf G_{10}\\
\mathbf 0
\end{matrix}\right) =
\left(\begin{matrix}
\mathbf \Phi\\
\mathbf B\\
\mathbf 0
\end{matrix}\right).
\end{equation}
\end{definition}
Since $\G_{tot}$ defines an MDS code, \code{0}, \code{1} and \code{10} (compare Section~\ref{sec:def} for the notations) are also MDS codes. 
We restrict $\varphi < k_1$ since otherwise all rows in $\mathbf G_1$ are rows of $\mathbf G_0$. 
Note that any rate $k/n$ in combination with any $k_1$ is feasible with this restriction since 
$k+1 \leq k+ k_1-\varphi \leq n$ and hence, we have only the trivial restriction  
%for a UM code with $k_1 =k$ and $\varphi = k-1$, the condition $k+k_1-\varphi \leq n$ 
%reduces to the trivial condition 
$k < n$.

\subsection{Calculation of Distances}\label{subsec:calcdist}
We calculate the extended row distance of the construction from Definition~\ref{def:pumcodearbrate} by cutting the semi-infinite generator matrix into parts. 
Each code block of length $n$ can be seen as a codeword of $\mathcal C_{\alpha}$ with minimum distance
\begin{equation*}
d_{\alpha} = d(\G_{\alpha})= d(\G_{tot})= n-k-k_1+\varphi +1.
\end{equation*}

However, due to the linear dependencies between the sub-generator matrices, a \emph{non-zero} information block can result in a \emph{zero} code block. 
The following lemma bounds the maximum number of such consecutive zero code blocks.\\[-4ex]
\begin{lemma}[Consecutive Zero Code Blocks]
The maximum number $\ell$ of zero code blocks $\mathbf c_j,\mathbf c_{j+1},\dots,\mathbf c_{j+\ell-1}$, which have no edge in common with the zero state, 
%(i.e., they result from non-zero information blocks)
is
\begin{equation*}
 \ell = \bigg \lceil \frac{\varphi}{k_1-\varphi}\bigg\rceil.
\end{equation*}

\end{lemma}
\begin{IEEEproof}
If $\varphi=0$, there is no zero code block obtained from a non-zero information block and $\ell = 0$. 

For $0 <\varphi< k_1$, let 
\begin{align*}
\mathbf i_{j-1}&=(\underbrace{i_0,\dots,i_{k_1-\varphi-1}}_{k_1-\varphi},\underbrace{0,\dots,0}_{\varphi}|\underbrace{i_{k_1},\dots,i_{k-1}}_{k-k_1})\\
\mathbf{i}_j&=(\underbrace{0,\dots,0}_{k_1-\varphi},\underbrace{i_0,\dots,i_{k_1-\varphi-1}}_{k_1-\varphi},\underbrace{0,\dots,0}_{\varphi-(k_1-\varphi)}|\underbrace{0,\dots,0}_{k-k_1})\\
&\vdots\\
\mathbf{i}_{j+\ell-2}&=(\hspace{-1ex}\underbrace{0,\dots,0}_{(\ell-1)(k_1-\varphi)}\hspace{-1ex},\underbrace{i_0,\dots,i_{k_1-\varphi-1}}_{k_1-\varphi},\hspace{-2ex}\underbrace{0,\dots,0}_{\varphi-(\ell-1)(k_1-\varphi)}\hspace{-1.5ex}|\underbrace{0,\dots,0}_{k-k_1})\\
\mathbf{i}_{j+\ell-1}&=(\underbrace{0,\dots,0}_{\ell(k_1-\varphi)},\underbrace{i_0,\dots,i_{\varphi-\ell(k_1-\varphi)-1}}_{\varphi-\ell(k_1-\varphi)}|\underbrace{0,\dots,0}_{k-k_1}).
\end{align*}
In the non-binary case, each second block $\mathbf{i}_j, \i_{j+2},\dots$ has to be multiplied by $-1$. 
Then,
\begin{equation*}
 \c_{j+h} = \i_{j+h-1} \cdot \mathbf G_1 + \i_{j+h} \cdot \mathbf G_0 = \mathbf 0, \quad \forall h=0,\dots,\ell-1.\\
% &\vdots\\
% \mathbf c_{j+\ell-1} &= \mathbf i_{j+\ell-2} \cdot \mathbf G_1 + \mathbf{i}_{j+\ell-1} \cdot \mathbf G_0 = \mathbf 0.
\end{equation*}
% For $0 <\varphi< k_1$, let $\i_{j-1}$ and $\i_j$ be as in \eqref{eq:info-cw-zero}. Then, $\c_j=\0$. 
% The next code block $\c_{j+1}=\0$ if $\i_{j+1}$ is again a shift of $\i_j$ to the right by $k_1-\varphi$ positions. 
In each step, we shift the information vector to the right by $k_1-\varphi$ positions, where this shift size is determined by the size of $\A$. 
Since $\mathbf \Phi$ has $\varphi$ rows, this right-shifting can be done $\left \lceil \varphi/(k_1-\varphi) \right\rceil$ times. 
%, where we denote the last information block by $\i_{j+\ell-1}$. 
We ceil the fraction since the last block $\i_{j+\ell-1}$ can contain less than $k_1-\varphi$ information symbols.
\end{IEEEproof}
Therefore, after $\ell$ zero code blocks there is at least one block of weight $d_{\alpha}$ and the slope can be lower bounded by:
\begin{equation}\label{eq:alpha}
 \alpha \geq \frac{d_{\alpha}}{\ell+1} = \frac{d_{\alpha}}{ \lceil \frac{\varphi}{k_1-\varphi}\rceil +1} = \frac{n-k-k_1+\varphi +1}{ \lceil \frac{k_1}{k_1-\varphi}\rceil}.
\end{equation}
The extended distances can be estimated as follows. \\[-4ex]
\begin{theorem}[Extended Distances]
The extended distances of order $j$ for the 
PUM code of Definition~\ref{def:pumcodearbrate} are:
%semi-infinite generator matrix given in \eqref{eq:winter-generator} are: 
\begin{align*}
 \dr{1} \geq & \ \drdes{1} = d_{01}, % = n-k + k_1+1,\\
  \ \ \dr{j} \geq \drdes{j} = d_0 + (j-2) \cdot \alpha + d_1, \ j>1,\\
%  &= 2\cdot (n-k+1) + (i-2)\cdot \frac{n-k-k_1+\varphi +1}{\ell+1},\\
 \dc{j} \geq & \ \dcdes{j}= d_0 + (j-1) \cdot \alpha, \ j >0,\\
%  &= n-k+1 + (i-1)\cdot \frac{n-k-k_1+\varphi +1}{\ell+1},\\
 \drc{j} \geq &\ \drcdes{j}=  (j-1) \cdot \alpha + d_1,\ j >0,
%   &= n-k+1 + (i-1)\cdot \frac{n-k-k_1+\varphi +1}{\ell+1}.\\
\end{align*}
with $d_{01}=n-k + k_1+1$, $d_0 = d_1 = n-k+1$ and $\alpha$ as in \eqref{eq:alpha} and 
$\drdes{j}$, $\dcdes{j}$ and $\drcdes{j}$ denote the designed extended distances.
\end{theorem}
\begin{IEEEproof}
For the calculation of the extended row distance, we start in the zero state, hence, the previous information is $\i_{0} = \0$. 
We obtain $d^r_1$ for an information block $\i_1=(0, \dots, 0, i^{(1)}_{k_1},\dots, i^{(1)}_{k-1})$, then $\c_1 \in \code{01}$. 
 The extended row distance of order $j$ follows from \eqref{eq:alpha} and a last information block
 $\i_{j}=(0, \dots, 0, i^{(j)}_{k_1},\dots, i^{(j)}_{k-1})$. The second-last block $\i_{j-1}$ is arbitrary and thus $\c_j = \mathbf i_j \cdot \mathbf G_0 + \mathbf i_{j-1} \cdot \mathbf G_1$
 is in $\code{1}$.
 
 The calculation of the extended column distance starts in the zero state, hence, $\i_{0} = \0$, but we end in any state, thus,  
 $\dc{1} \geq d_0$. For higher orders, each other block is in $\code{\alpha}$.
 
 The reverse extended column distances considers all code blocks starting in any state, hence there is no restriction on $\i_0, \i_1$ and 
 $\c_1 \in \code{\alpha}$. In order to end in the zero state, $\i_{j}=(0, \dots, 0, i^{(j)}_{k_1},\dots, i^{(j)}_{k-1})$ and as for the extended row distance 
 $\c_j \in \code{1}$.
\end{IEEEproof}

The free distance is then the minimum, i.e., 
\begin{equation*}
\dfree \geq \min_{\substack{i=1,2,\dots}} \lbrace \dr{i}\rbrace = \min \lbrace n-k + k_1+1,2\cdot (n-k+1)  \rbrace.
\end{equation*}
Note that if $\dfree = n-k + k_1+1$, then the free distance is optimal since the upper bound is achieved \cite{Pollara_FiniteStateCodes}.

%%%%%%%%%%%%%%%%%%%%%%%%%%%%%%%%%%%%%%%%%%%%%%%%%%%%%%%%%%%%%%%%%%%%%%%%%%%%%%%%%%%%%%%%%%%%%%%%%%%%%%%%%%%%%%%%%%%%%%%%%%%%%%%%%%%%%%%%%%%%%%%%%%%%%%%%%%%%%%%%%%%%%%%%%%%%%%%%%%%%%%%%%%%%%%%%%%%%%%%%%%%%%%%%%%%%
\section{BMD Decoding Algorithm}\label{sec:decoding}
%%%%%%%%%%%%%%%%%%%%%%%%%%%%%%%%%%%%%%%%%%%%%%%%%%%%%%%%%%%%%%%%%%%%%%%%%%%%%%%%%%%%%%%%%%%%%%%%%%%%%%%%%%%%%%%%%%%%%%%%%%%%%%%%%%%%%%%%%%%%%%%%%%%%%%%%%%%%%%%%%%%%%%%%%%%%%%%%%%%%%%%%%%%%%%%%%%%%%%%%%%%%%%%%%%%%

\subsection{BMD Condition and Idea}
% Dettmar describes in \cite{DettmarSorger_BMDofUM} a BMD decoding algorithm for errors up to \drhalf.
% Therefore, we define a BMD Decoder for convolutional codes.
% \begin{definition}[BMD Decoder for Convolutional Codes \cite{DettmarSorger_BMDofUM}]
Let the received sequence $\r = \c+\e=(\r_0,\r_1,\dots,\r_{N-1})$ be given, where $\r_h=\c_h+\e_h$, $h=0,\dots,N-1$ is in $\F^n$, 
$\c = (\c_0, \c_1,\dots,\c_{N-1})$ is a codeword of the (terminated) PUM code as in Definition~\ref{def:pumcodearbrate} and 
$\e_h$ is an error block of Hamming weight $\wt(\e_h)$. 
%In this paper, we give a so-called BMD decoding algorithm for PUM codes of arbitrary rate.
%, where a
A BMD decoder for convolutional codes is defined as follows.\\[-5ex]
\begin{definition}[BMD Decoder for Convolutional Codes \cite{DettmarSorger_BMDofUM}]
A BMD decoder for convolutional codes guarantees 
%the correct decoding of a received vector $\mathbf r=\mathbf c+\mathbf e$ of length $n$ if the error vector $\mathbf e$ satisfies
to find the Maximum Likelihood (ML) path as long as
\begin{equation}\label{eq:bmdcond}
\sum\limits_{h=j}^{j+i-1}\wt (\e_{h}) <\frac{\drdes{i}}{2} %, \; \forall j = 0,\dots,N-1, i=1,\dots,N-j. %&0&\leq t+i \leq n\ .
%\wt\left(\mathbf{e}_{t+0}\ \mathbf{e}_{t+1}\ \dots\ \mathbf{e}_{t+i}\right)&<\frac{\dr{i}}{2}, &0&\leq t+i \leq n\ .
\end{equation}
holds for all $j = 0,\dots,N-1$ and $i=1,\dots,N-j$.
\end{definition} 
% where $\wt(\e_h)$ is the Hamming weight of the error block $h$. 
% A similar condition can be given for \emph{one} block of the PUM code.
Algorithm~\ref{alg:pum} shows the basic principle
of our generalization of the Dettmar--Sorger algorithm to arbitrary rate. %the construction from Definition~\ref{def:pumcodearbrate}. 

\begin{figure}[htb]
\begin{center}
\includegraphics[width=0.75\columnwidth]{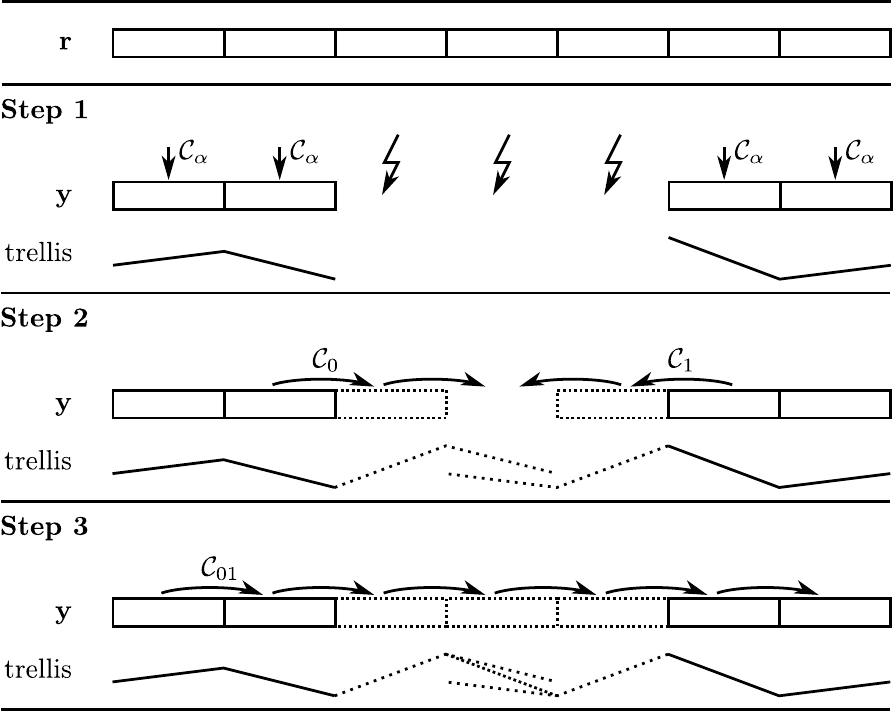}
\end{center}
\caption{Example of the decoding algorithm for $\ell=1$, where the three first steps of 
Algorithm~\ref{alg:pum} for the received sequence $\r$ are illustrated.}
\label{fig:reduced_trellis}
\end{figure}

% \begin{algorithm}
\printalgo{\caption{Arbitrary-Rate Decoder for PUM codes}
\label{alg:pum}
\dontprintsemicolon
\SetVline
\BlankLine
\linesnumbered
% \DontPrintSemicolon
% \SetAlgoLined
% \LinesNumbered
\KwIn{Received sequence $\mathbf{r}$ of length $N\cdot n$}
\BlankLine
Decode block $\r_0$ with \BMD{\code{0}},\phantom{teeeeeeeeeeeeeeeeeeest}
decode blocks $\mathbf{r}_j$ for $j=1,\dots,N-2$ with \BMD{\code{\alpha}},
decode block $\r_{N-1}$ with \BMD{\code{1}},\phantom{teeeeeeeeeeeeeeeest}
calculate $\mathbf{i}_j$ if $\ell+1$ consecutive blocks were decoded successfully and assign metric as in \eqref{eq:metric_step1}\;
\BlankLine
From all found blocks $\mathbf{i}_j$, decode $\ell_F^{(j)}$ steps forwards with \BMD{\code{0}} and $\ell_B^{(j)}$ steps backwards with \BMD{\code{1}}\;
\BlankLine
From all found blocks $\mathbf{i}_j$, decode next block with \BMD{\code{01}} and assign metric as in \eqref{eq:defmetric_step3}\; %and calculate the metric of the found edges\;
\BlankLine
Search the complete path of smallest weight with the Viterbi algorithm\;
\BlankLine
\KwOut{Information sequence $\mathbf{i}$ of length $(N-1)\cdot k$}
}
% \begin{small}
% \input{algos/bmd_decoding.tex}
% \end{small}
% \end{algorithm}
The main idea of the algorithm is to take advantage of the efficient 
BMD block decoders for \code{\alpha}, \code{0}, \code{1} and \code{01}. 
With the results of the block decoders, we build a reduced trellis and finally 
use the Viterbi algorithm to find the ML path. Since this trellis has only very few edges, 
the overall decoding complexity is only cubic in the length. 
Figure~\ref{fig:reduced_trellis} illustrates the decoding principle for $\ell=1$.

Since each code block of the PUM code of length $n$ is a codeword of the block code \code{\alpha}, 
the first step of the algorithm is decoding with \BMD{\code{\alpha}}. % since each code block is a codeword of \code{\alpha}.
Due to the termination, the first and the last block can be decoded with \BMD{\code{0}}, respectively \BMD{\code{1}}.
The decoding result of \BMD{\code{\alpha}} is $\bar{\c}_j$. Assume it is correct, then $\bar{\c}_j={\c}_j=\i_j\G_0 + \i_{j-1}^{[k_1]} \G_{10}$, where $\i_{j-1}^{[k_1]}=(i_0^{(j-1)}, \dots, i_{k_1-1}^{(j-1)})$ is a part of the previous information block. 
Now, we want to reconstruct the information 
$\i_j = (i_0^{(j)}, \dots,  i_{k-1}^{(j)})$ and $\i_{j-1}^{[k_1]}$. 
For this, we need $\ell+1$ consecutive decoded code blocks since the linear dependencies ``spread'' to the next $\ell$ blocks 
as shown in Example~\ref{ex:reconst_info}. \\[-4ex]
\begin{example}[Reconstructing the Information]\label{ex:reconst_info}
%Let us illustrate the reconstruction for 
Let $\varphi = 2/3 k_1$, where $\ell=2$ and $\mathbf \Phi$ has 
twice as much rows as $\A$. Assume, we have decoded $\c_0$, $\c_1$ and $\c_2$ and
we want to reconstruct $\i_1$. Decompose $\i_0, \i_1, \i_2$ into: 
$\i_j = (\i_j^{[1]}\; |\; \i_j^{[2]}\; |\;\i_j^{[3]}\; |\;\i_j^{[4]} )$ for $j=0,1,2$, where the first three sub-blocks
have length $k_1-\varphi$ and the last
$k-k_1$. Then,
\begin{equation*}
\c_1 = %\underbrace{
(\i_1^{[1]}\; |\; \i_1^{[2]} +\i_0^{[1]} \; |\;\i_1^{[3]}+ \i_0^{[2]}\; |\;\i_1^{[4]}   \; |\;\i_0^{[3]} )
%}_{ \defeq \ \widehat{\i}_1}
\begin{pmatrix}
\A \\
\mathbf{\Phi}_1\\
\mathbf{\Phi}_2\\
\G_{01}\\
\B
\end{pmatrix}
\defeq \widehat{\i}_1 \cdot \G_{tot},
\end{equation*}
where $\mathbf \Phi = \left(\begin{smallmatrix}
\mathbf \Phi_1\\\mathbf \Phi_2
\end{smallmatrix}\right)$ and $\mathbf \Phi_1$, $\mathbf \Phi_2$ have $k_1-\varphi$ rows. 
Since we know $\c_1$ and $\G_{tot}$ defines an MDS code, we can reconstruct the vector $\widehat{\i}_1$. 
This directly gives us $\i_1^{[1]}$ and $\i_1^{[4]} $. This can be done
in the same way for $\c_0$ and we also directly obtain (among others) $\i_0^{[1]}$. 
To obtain $\i_1^{[2]}$, we substract $\i_0^{[1]}$ from the known sum $\i_1^{[2]} +\i_0^{[1]}$.
For $\c_2$, this reconstruction provides $\i_1^{[3]} $ and we have the whole
$\i_1$. 
This principle also gives us $\i_{0}^{[k_1]} = (\i_0^{[1]}\; |\; \i_0^{[2]}\; |\;\i_0^{[3]})$. 
This is why $\ell+1$ consecutive decoded blocks are necessary to reconstruct an information block. 
Note that it does not matter if the other decoded blocks precede or succeed the wanted information, this 
principle works the same way.
\end{example}

After this decoding and reconstruction, we build an edge in a reduced trellis for each block with the metric: 
\begin{equation}\label{eq:metric_step1}
m_{j}=\begin{cases}
                  \ \wt(\mathbf r_j-\bar{\mathbf c}_j)\quad \ \text{ if Step~1 finds }\bar{\mathbf c}_j\text{ and }\bar{\i}_j,\\ %\BMD{\code{\alpha}} decodes } \mathbf c_j^{(\alpha)}\text{ and }\i_j \text{could be reconstructed}\\
		  \ \lfloor(d_\alpha+1)/2\rfloor\ \ \text{ else.}
                 \end{cases}
\end{equation}\\[-5ex]
\begin{remark}\label{rem:reason_ell}
 The error of minimum weight causing a sequence of non-reconstructed information blocks 
 in Step~1 is as follows: %decoding failures in Step~1 for a whole sequence is as follows:
 \begin{equation*}
  (\underbrace{0, \dots, 0, \times}_{\ell+1 \text{ blocks}} \ |\underbrace{ \ 0, \dots, 0, \times \ }_{\ell+1\text{ blocks}}|\ \dots \ |\underbrace{\ 0, \dots,0, \times\ }_{\ell+1\text{ blocks}} |\underbrace{ \ 0,\dots, 0}_{\ell\text{ blocks}}),
 \end{equation*}
where the $\times$ marks blocks with at least $\da/2$ errors. 
Also the information of the error-free blocks cannot be reconstructed, since we need $\ell+1$ consecutive decoded blocks. 
The last $\ell$ error-free blocks are the reason why we substract $\ell$ in the definitions of $\ell_F^{(j)}$ and $L_F^{(j)}$. 
This corresponds to $\ell$ additional decoding steps in forward direction. 
The (minimum) average weight in a sequence of non-reconstructed information blocks (without the last $\ell$ blocks) is 
therefore $\da/(2(\ell+1))$. 
\end{remark}

% Assume, a sequence of received blocks $\mathbf r$ occurs, where some blocks cannot be decoded with \code{\alpha}. 
%with more than \dahalf errors every $\ell+1$th block which cannot be decoded with \code{\alpha}.
% These blocks are located between blocks with less than $\dahalf$ errors, which can be decoded successfully with \code{\alpha} and are used to construct a reduced trellis.
Assume, in Step~1, we decoded $\c_j$ and reconstructed $\i_j $ and a part of the previous information $\i_{j-1}^{[k_1]}$, then we calculate:
\begin{align}
 &\r_{j+1} - (i_0^{(j)}, \dots, i_{k_1-1}^{(j)}) \cdot \G_{10} = \i_{j+1} \cdot \G_0 + \e_{j+1}\nonumber\\
 &\r_{j-1} - (i_0^{(j-1)}, \dots, i_{k_1-1}^{(j-1)}) \cdot \G_{00}\label{eq:substract_decoding}\\ & \hspace{2ex}= (i_{k_1}^{(j-1)}, \dots, i_{k-1}^{(j-1)}|i_0^{(j-2)}, \dots, i_{k_1-1}^{(j-2)}) \cdot \left(\begin{matrix}\mathbf G_{01}\\\mathbf G_{10}\end{matrix}\right)\nonumber + \e_{j-1}.
\end{align}
Hence, as a second step, 
%we use the information gained from the decoded blocks 
we decode $\ell_F^{(j)}$ blocks forward with \BMD{\code{0}} respectively $\ell_B^{(j)}$ blocks backward in \BMD{\code{1}}. 
These codes have higher minimum distances than $\da$ and close (most of) the gaps between two sequences of correctly decoded blocks in \code{\alpha}.
The values $\ell_F^{(j)}$ and $\ell_B^{(j)}$ are defined by:
\begin{align}
\ell_F^{(j)}&=\min_{i=1,2,\dots}\Big(i\Big\arrowvert\sum\limits_{h=1}^{i-\ell}\frac{d_\alpha-m_{j+h}}{\ell+1}\geq \frac{\dcdes{i}}{2}\Big),\label{eq:deflf}\\
\ell_B^{(j)}&=\min_{i=1,2,\dots}\Big(i\Big\arrowvert\sum\limits_{h=1}^i\frac{d_\alpha-m_{j-h}}{\ell+1}\geq \frac{\drcdes{i}}{2}\Big).\label{eq:deflb}
\end{align}
Lemma~\ref{lem:gapsize1} in Section~\ref{subsec:proofalgo} proves that after Step~2, the size of the 
gap between two correctly reconstructed blocks is at most one block. 
 
For Step~3, assume we know $\i_{j-1}^{[k_1]}=(i_0^{(j-1)}, \dots, i_{k_1-1}^{(j-1)})$ from Step~1 and $\i_{j-2}$ from Step~1 or 2, then similar to \eqref{eq:substract_decoding}:
\begin{align*}
&\r_{j-1} - (i_0^{(j-1)}, \dots, i_{k_1-1}^{(j-1)}) \cdot \G_{00} - (i_0^{(j-2)}, \dots, i_{k_1-1}^{(j-2)}) \cdot \mathbf G_{01} \nonumber\\ 
& \hspace{5ex}= (i_{k_1}^{(j-1)}, \dots, i_{k-1}^{(j-1)}) \cdot \mathbf G_{01}+ \e_{j-1}\nonumber,
\end{align*}
which shows that we can
use \BMD{\code{01}} to close the remaining gap at $j-1$. 
After Step~3, assign as metric to each edge
\begin{equation}\label{eq:defmetric_step3}
m_{j}=\begin{cases}
                  \ \wt(\mathbf r_j-\bar{\mathbf c}_j)\qquad  \parbox[t]{\linewidth}{\begin{small}
                  if \BMD{\code{0}}, \BMD{\code{1}} or \BMD{\code{01}}\\ is successful and $\bar{\mathbf i}_j$ is reconstructed,
                  \end{small}}\\[3ex]
		   \lfloor(d_{01}+1)/2\rfloor \quad \ \; \text{\begin{small}else\end{small}},
                 \end{cases}     
\end{equation}
where again $\bar{\mathbf c}_j$ denotes the result of a successful decoding. 
Note that there can be more than one edge in the reduced trellis at depth $j$.

% Finally, if \eqref{eq:bmdcond} holds, the correct path is found with the Viterbi algorithm for which the metric of edges is given in Definition \ref{def:metric}.
Finally, we use the Viterbi algorithm to search the ML path in this reduced trellis. 
As in \cite{DettmarSorger_BMDofUM}, we use $m_{j}$ as edge metric and the sum over different edges as path metric. 
%Theorem~\ref{the:winter_correct} proves that if \eqref{eq:bmdcond} is fulfilled, we will find the ML path. 

Section~\ref{subsec:proofalgo} proves that if \eqref{eq:bmdcond} is fulfilled, after Steps~1--3, all gaps are closed
and Algorithm~\ref{alg:pum} finds the ML path.
% \begin{definition}[Metric]\label{def:metric}
% The metric of an edge $e(\mathbf{s}_{t-1}^{(j)},\mathbf{s}_{t}^{(i)})$ between two nodes $\mathbf{s}_{t-1}^{(j)}\in S_{t-1}$ and $\mathbf{s}_{t}^{(i)}\in S_t$ is defined as
% \begin{align*}
% \m{\mathbf{i}_{t+1}^{(i)}}=\m{\mathbf{i}_{t}^{(j)}}+\dist{\mathbf{e}_{t+1}}\ ,
% \end{align*}
% where $\dist{\mathbf{e}_{t+1}}$ is the Hamming weight of the error $\mathbf{e}_{t+1}$, $\mathbf{i}_{t+1}^{(i)},\mathbf{i}_{t}^{(j)}$ are the corresponding information blocks and $\mathbf{s}_{t-1}^{(j)}\in S_{t-1},\mathbf{s}_{t}^{(i)}\in S_t$ are the syndrome edges of the syndrome trellis.
% \end{definition}
%In the blocks in the middle there occurred more than \drhalf errors which can be corrected by forward and backward decoding with \code{0} and \code{1}.
% The adapted algorithm is outlined as Algorithm \ref{alg:pum}.
%In Section~\ref{subsec:proofalgo}, we prove that Algorithm~\ref{alg:pum} finds the ML path if \eqref{eq:bmdcond} is fulfilled. 
It is a generalization of the Dettmar--Sorger algorithm to arbitrary rates, which
results in linear dependencies between the submatrices of the PUM code (see Definition~\ref{def:pumcodearbrate}).
This requires several non-trivial modifications of the algorithm. Namely these are:
the reconstruction of the information requires $\ell +1$ consecutive code blocks (see Example~\ref{ex:reconst_info}), 
the path extensions \eqref{eq:deflf}, \eqref{eq:deflb} have to be prolonged and
the assigned metric has to be adapted appropriately \eqref{eq:metric_step1}, \eqref{eq:defmetric_step3}
since the smallest error causing a non-reconstructable sequence is generalized as in Remark~\ref{rem:reason_ell}.\\[-5ex]
%in order 
%to guarantee that we find the ML path if \eqref{eq:bmdcond} is fulfilled.
\subsection{Proof of Correctness}\label{subsec:proofalgo}
%We prove that we find the ML path if \eqref{eq:bmdcond} is satisfied.
In this subsection, we prove that Algorithm~\ref{alg:pum} finds the ML path if \eqref{eq:bmdcond} is fulfilled. 
For this purpose, Lemma~\ref{lem:gapsizesmall} shows that the size of the gaps after Step~1 is not too big and
in Lemma~\ref{lem:gapsize1} we prove that after Step~2, the gap size is at most one block. Finally, Theorem~\ref{the:winter_correct} shows 
that we can close this gap and that the ML path is in the reduced trellis. Then, the Viterbi algorithm will find it. 
The complexity of the decoding algorithm is stated in Theorem~\ref{theo:complexity}.\\[-4ex]
\begin{lemma}\label{lem:gapsizesmall}
The length of any gap between two correct reconstructions in Step~1, $\i_j$, $\i_{j+i}$, is less than $\min(L_F^{(j)},L_B^{(j+i)})$ if \eqref{eq:bmdcond} holds, with\\[-3ex]
\begin{align*}
L_F^{(j)}&=\min_{i=1,2,\dots}\Big(i\Big\arrowvert\sum\limits_{h=1}^{i-\ell} \frac{d_\alpha-m_{j+h}}{\ell+1}\geq \frac{\drdes{i}}{2}\Big),\\
L_B^{(j)}&=\min_{i=1,2,\dots}\Big(i\Big\arrowvert\sum\limits_{h=1}^i \frac{d_\alpha-m_{j-h}}{\ell+1}\geq \frac{\drdes{i}}{2}\Big).
\end{align*}
\end{lemma}
\begin{IEEEproof}
% This is an estimation of the smallest weight an alternative path to the ML path would have.
% Block--BMD decoding of \code{\alpha} fails in 
Step~1 fails if there occur %more than $\dahalf$ errors in every $(\ell+1)$-th block.
at least $\da/2$ errors in every $(\ell+1)$-th block, followed by $\ell$ correct ones (compare Remark~\ref{rem:reason_ell}).
Assume there is a gap of at least $L_F^{(j)}$ blocks after Step~1. 
Then,
\begin{equation*}
 \sum\limits_{h=1}^{L_F^{(j)}}\wt\left(\mathbf{e}_h\right)
 \geq \!\!\!\sum\limits_{h=1}^{L_F^{(j)}-\ell}\!\!\frac{\da}{2(\ell+1)}
\geq \!\!\!\sum\limits_{h=1}^{L_F^{(j)}-\ell}\!\frac{(d_\alpha-m_{j+h})}{\ell+1}\geq
\!\frac{\drdes{L_F^{(j)}}}{2},
\end{equation*}
contradicting \eqref{eq:bmdcond}. We prove this similarly for $L_B^{(j)}$ without substracting $\ell$ in the limit of the sum,
since we directly start left of the $\ell$ correct blocks on the right. 
Therefore, the gap size is less than $\min(L_F^{(i)},L_B^{(i)})$.
\end{IEEEproof}
\vspace{-1ex}
\begin{lemma}\label{lem:gapsize1}
Let $\mathbf{i}_j$ and $\mathbf{i}_{j+i}$ be reconstructed in Step~1. Let Step~2 decode $\ell_F^{(j)}$ blocks in forward and $\ell_B^{(j+i)}$ blocks in backward direction (see \eqref{eq:deflf}, \eqref{eq:deflb}).
Then, except for at most one block, the ML path is in the reduced trellis if \eqref{eq:bmdcond} holds. 
\end{lemma}
\begin{IEEEproof}
First, we prove that the ML path is in the reduced trellis if \eqref{eq:bmdcond} holds and in each block
less than $\min \lbrace d_0/2, d_1/2 \rbrace$
errors occurred. 
%If there are  
%less than $\min \lbrace d_0/2, d_1/2 \rbrace$
%errors in a block, 
In this case, 
\BMD{\code{0}} and \BMD{\code{1}} will always yield the correct decision.
The ML path is in the reduced trellis if $\ell_F^{(j)}+\ell_B^{(j+i)}\geq i-1$, since the gap is then closed.
Assume that $\ell_F^{(j)}+\ell_B^{(j+i)}<i-1$ and at least $\da/2$ errors occur in every $(\ell+1)$-th block in the gap, 
since Step~1 was not successful (compare Remark~\ref{rem:reason_ell}).
Then,
\begin{align*}
\sum\limits_{h=1}^{i-1}\!\wt\left(\mathbf{e}_{j+h}\right)& \!\geq\! \frac{\dcdes{\ell_F^{(j)}}}{2}\!+\!\frac{\drcdes{\ell_B^{(j+i)}}}{2}%\\%+\frac{d_\alpha}{2}+\\
%&+(i-1-\ell_F^{(t)}-\ell_B^{(j+i)})\cdot\frac{d_\alpha}{2(\ell+1)}=\\
+\frac{(i-1-\ell_F^{(t)}-\ell_B^{(j+i)})d_\alpha}{2(\ell+1)}=\\
%&=d_{0}+(\ell_F^{(t)}-1+j-1-\ell_F^{(t)}-\ell_B^{(t+j)}+\ell_B^{(t+j)}-1)\cdot\frac{d_\alpha}{2(\ell+1)}=\\
&=\frac{d_{0}}{2}+\left(i-3\right)\cdot\frac{d_\alpha}{2(\ell+1)}+\frac{d_{1}}{2}=\frac{\drdes{i-1}}{2},
\end{align*}
which is a contradiction to \eqref{eq:bmdcond}.

Second, we prove that at most one error block $\mathbf{e}_h$, $j <h <j+i$ has weight at least $d_0/2$ or $d_1/2$.
To fail in Step~1, there are at least $\da/2$ errors in every $(\ell+1)$-th block.
If two error blocks have weight at least $d_0/2=d_1/2$, then %more than $\geq\lceil\frac{d_0+1}{2}\rceil$, there would follow
\begin{align*}
\sum\limits_{h=1}^{i-1}\wt\left(\mathbf{e}_{j+i}\right)\geq2\cdot\frac{d_0}{2}+\frac{i-3}{\ell+1}\cdot \frac{d_\alpha}{2}\geq \frac{\drdes{i-1}}{2},
\end{align*}
in contradiction to \eqref{eq:bmdcond}. 
Thus, the ML path is in the reduced trellis except for a gap of one block.
\end{IEEEproof}
\vspace{-1ex}
\begin{theorem}\label{the:winter_correct}
If \eqref{eq:bmdcond} holds, the ML path is in the reduced trellis.
\end{theorem}
\begin{IEEEproof}
Lemma~\ref{lem:gapsize1} guarantees that after Step~2, the gap length is at most one block. This gap can be closed in Step~3 with \code{01}, which is always able to find the correct solution since $d_{01} \geq \drdes{1}=\dfree$. %it can correct more errors than half the minimum distance $\frac{\dfree}{2}$.
\end{IEEEproof}

\subsection{Decoding of a Single Block}
Similar to \cite{DettmarSorger_BMDofUM}, we give a weaker BMD condition to guarantee ML decoding of a \emph{single} block.
This condition shows how fast the algorithm returns to the ML path after a sequence where \eqref{eq:bmdcond} is not fulfilled.
% \begin{definition}[Decoding of Single Blocks \cite{DettmarSorger_BMDofUM}]
A BMD decoder for convolutional codes guarantees the correct decoding of a block $\mathbf{r}_j$ of a received sequence $\mathbf{r}=\mathbf{c}+\mathbf{e}$ if the error $\mathbf{e}$ satisfies
\begin{align}\label{eq:bmdblock}
% \sum\limits_{i=k}^{k+j-1}\wt\left(\mathbf{e}_i\right)&<\frac{\dr{t-k+1}}{2}, &k&\leq t,\ j\geq t-k+1\ .
\sum\limits_{h=k}^{k+i-1}\wt\left(\mathbf{e}_h\right)&<\frac{\drdes{i}}{2}, \ \forall i,k \text{ with } k\leq j \leq j+i-1.
\end{align}
% \end{definition}

% To estimate the rest of the trellis, erasure nodes are introduced with the following metric:
To %ensure that the distance between received and decoded sequence is less than $\drdes{i} /2$, 
guarantee \eqref{eq:bmdblock} for a certain block if \eqref{eq:bmdcond} is not fulfilled for the whole sequence, 
we introduce an \emph{erasure node} in each step $j$ as in \cite{DettmarSorger_PUMBCH}, representing all nodes which are not in the reduced trellis. 
Let $\en{j}$, $\en{j-1}$ denote erasure nodes at time $j$, $j-1$ and let $s_j$, $s_{j-1}$ be nodes found by BMD decoding in Steps~1 and 2. 
Let $t_F,t_B$ denote the minimum number of errors of any edge starting from $s_{j-1}$ and $s_j$ in forward, respectively backward direction. 
$t_{\alpha}$ denotes the minimum number errors of any edge between nodes at time $j-1$ and $j$. 
We set the metric of the connections with the erasure nodes as follows.

\begin{small}
\begin{center}
% \begin{table}
% \caption{Edge metric definition including erasure nodes}
\setlength{\extrarowheight}{1.5ex}
\begin{tabular}[]{c|l}\label{tab:winter_en}
%$S_{t-1}$ & $S_{t}$ & $\m{e_t}$\\
Connect&Metric\\
% $j-1$&$j$&\\
\hline
$s_{j-1}$, $\en{j}$ & $\m{\en{j}}=m(s_{j-1})+\frac{\max{(\lfloor(d_0+1)/2\rfloor, \; d_0-t_F)}}{\ell+1}$\\
\hline
$\en{j-1}$, $s_j$ & $m(s_j)=m(\en{j-1})+ \frac{\max{\left(\left\lfloor(d_1+1)/2\right\rfloor,\; d_1-t_B\right)}}{\ell+1}$\\
\hline
$\en{j-1}$, $\en{j}$ & $\m{\en{j}}=m(\en{j-1}) +$\\ %[-0.5ex]
&$\!\!+\frac{1}{\ell+1}\!\cdot\!
\begin{cases}
(d_\alpha-t_{\alpha})\quad \text{ if }\exists\text{ an edge between } s_{j-1},s_j\!\!\!\!\!\!\!\!\!\!\\[-0.2ex]
\lfloor(d_\alpha+1)/2\rfloor\text{, else.}
\end{cases}$\\
%$\left\lbrace
%\begin{matrix}
%&(d_\alpha-t_{\alpha})\text{ if }\exists\text{ an edge between } s_{j-1},s_j\\ %\!\!\!\!\!\!\!\!\!\!\\[-0.2ex]
%&\lfloor(d_\alpha+1)/2\rfloor\text{, else.}
%\end{matrix} \right.$\\%$
\end{tabular}
% \end{table}
%\caption{Metrics of Erasure Nodes}
%\label{tab:en}
\end{center}
\end{small}

\begin{theorem}
If \eqref{eq:bmdblock} holds for $\mathbf{r}_j$, the Viterbi algorithm for the reduced trellis with erasure nodes finds the correct block $\c_j$.
\end{theorem}
\begin{IEEEproof}
The metric of the erasure nodes is always at least $\drdes{i}/2$. All nodes of a state are connected with the erasure nodes of the previous and the next state.
As soon as \eqref{eq:bmdblock} is fulfilled, the metric of a correct edge is better than all other edges and the ML path will be chosen.
\end{IEEEproof}
% These proofs contain no restrictions for UM code constructions where $k_1=k$. In fact, Step 3 is obsolete since according to \ref{lem:wintergap}, the remaining gap after Step 2 is only 1 block long if \eqref{eq:bmdcond} holds. This information is encoded into the neighbour blocks so that the ML path is already in the reduced trellis. The erasure nodes do not have any restrictions as well.

\subsection{Complexity Analysis}
% An upper bound for the decoding complexity will be given.
The complexity is determined by the complexity of the BMD block decoders, which are all in the order $\mathcal O(n^2)$, 
if the construction is based on RS codes of length $n$. 
%which will be denoted with $\comp_\alpha$, $\comp_0$, $\comp_1$ and $\comp_{01}$ and the number of decodings used by the algorithm.
%We also introduce $\comp_B=\max\{\comp_\alpha,\comp_0,\comp_1,\comp_{01}\}$. We use RS codes so that $\comp_B=\order{n^2}$.
% Due to the steps we obtain:
% \begin{align*}
% \comp_{(P)UM}&\leq \comp_\alpha+(\ell_F+\ell)\comp_0+\ell_B\comp_1+(N_S-1)\comp_{01}\leq\\
% &\leq (N_S+\ell_F+\ell_B+\ell)\order{n^2}\ ,
% \end{align*}
% where $N_S$ is the number of nodes in a state without the erasure node.
% Successful decodings with \code{0} and \code{1} will not be repeated and since we know that the number of nodes in a state is $N_S\leq \ell_F+\ell_B+\ell$, we obtain
% \begin{align*}
% \comp_{(P)UM}\leq2 N_S \order{n^2}\ .
% \end{align*}

Similar as Dettmar and Sorger \cite{DettmarSorger_BMDofUM}, we can give the following bound on the complexity.
Due to space restrictions, the proof is omitted here.\\[-4ex]
\begin{theorem}\label{theo:complexity}
Let \code{} be a PUM code as in Definition~\ref{def:pumcodearbrate}, where $\G_{tot}$ is the generator matrix of an RS code.
Then, the decoding complexity of Algorithm~\ref{alg:pum} of one block is upper bounded by
\begin{equation*} %\label{eq:com_bound}
\comp_{PUM}\leq\order{(\ell+1) d_\alpha n^2} \sim \order{(\ell+1) n^3}.
\end{equation*}
\end{theorem}

\section{Conclusion}\label{sec:concl}
We presented a construction of PUM codes of arbitrary rate and provided and proved an
efficient decoding algorithm. 
The algorithm corrects all error patterns up to half the designed extended row distance, where the complexity is cubic in the length of a block.
For $\ell = 0$, the Dettmar--Sorger algorithm \cite{DettmarSorger_BMDofUM} is a special case of Algorithm~\ref{alg:pum}.
\section*{Acknowledgment}
The authors thank Alexander Zeh and Vladimir Sidorenko for the valuable discussions.

\bibliographystyle{IEEEtran}
\bibliography{antoniawachter}
\end{document}